\documentclass[a4paper]{jpconf}
\usepackage{graphicx}
\usepackage{lineno}
\usepackage{rotating}
\usepackage{multirow}
\usepackage{graphicx,epsfig}
\usepackage{subfigure}
\usepackage{mathrsfs}
\usepackage{amssymb}
\usepackage{cite}
\usepackage{xspace}
\usepackage{graphicx}
\usepackage{color}

\newcommand{\ruu}{\ensuremath{^{\circ \circ}}\xspace}
\newcommand{\rcu}{\ensuremath{^{\bullet \circ}}\xspace}
\newcommand{\rcc}{\ensuremath{^{\bullet \bullet}}\xspace}
\newcommand{\rcp}{\ensuremath{^{\bullet \ast}}\xspace}
\newcommand{\rpp}{\ensuremath{^{\ast \ast}}\xspace}
\newcommand{\rnu}{\ensuremath{^{\times \circ}}\xspace}

\newcommand{\rnp}{\ensuremath{^{\times \ast}}\xspace}

\newcommand{\ttbar}{\ensuremath{t \bar t}\xspace}
\newcommand{\met}{\ensuremath{E_T^{\rm miss}}}
\newcommand{\mt}{\ensuremath{m_{\rm top}}}

\begin{document}
\title{Review of recent top-quark LHC combinations}

\author{Giorgio Cortiana, on behalf of the ATLAS and CMS
  collaborations within the TOPLHCWG}

\address{Max-Planck-Institut f{\"u}r Physik, F{\"o}hringer Ring 6, D-80805 M{\"u}nchen, Germany.}

\ead{cortiana@mpp.mpg.de}

\begin{abstract}
  A review of recent combinations of top-quark measurements performed
  at the LHC, by the ATLAS and CMS collaborations, is
  provided. The typical uncertainty categorisations, and their assumed
  correlation patterns are presented, together with  the results of the
  combinations of the top-quark pair and single top-quark production
  cross sections, the top-quark mass, as well as of the $W$ boson
  polarisation and the charge asymmetry in \ttbar\ events.
\end{abstract}

\section{Introduction}

The TOPLHCWG~\cite{TOPLHCWG}, formed in 2011, constitutes a forum for
the study of the experimental and theoretical uncertainties affecting
the measurements of the top-quark properties at the LHC. The main
objective of the working group consists in the combination of the
results of the ATLAS and CMS collaborations and their presentation
allowing clear theoretical interpretations.
It is structured in several sub-groups to perform specific
combination tasks and to conduct topical discussions involving
analysers and experts from both collaborations.
The scientific output of the working group consists of combinations of
individual top-quark measurements and of various sets of
recommendations aimed at refining and harmonising the statistical and
systematic uncertainty treatment in current and future measurements.
In this document a review of the recent LHC combination results of the
top-quark properties will be presented.
The combinations are performed using the Best Linear Unbiased Estimate
(BLUE) method~\cite{BLUE1, BLUE2}.  BLUE determines the coefficients
(weights) to be used in a linear combination of the input measurements
by minimising the total uncertainty of the combined result.  In the
algorithm, assuming that individual measurements are unbiased and that
all uncertainties are distributed according to Gaussian probability
density functions, both statistical and systematic uncertainties, and
their correlations, are taken into account.
%
%
Input to all combinations are the individual results with a detailed
breakdown of the uncertainties as well as the assumed correlations
between individual sources. The tasks of each combination effort is to
determine a mapping between corresponding uncertainties sources and to
understand the correlations in each of the categories across different
experiments.

\section{Main systematic uncertainty categories}

Due to the large top-quark samples available at the LHC,
the precision of the measurements is typically limited by systematic
uncertainties. As detailed in the following, these can be grouped into three main categories:
theoretical, experimental and background or luminosity related
uncertainties. In general all systematic uncertainties of
measurements performed within the same experiment and during the same
data taking period are considered fully correlated.

\subsection{Theory based uncertainties}
Theory based uncertainties are related to the simulation of top-quark
signal events, to the event modelling and the description of the hard
scattering environment. Choices to be made in the signal simulation
are the proton distribution functions (PDF), the Monte Carlo
generator (MC) and the hadronisation model.  On the event modelling
side, important ingredients are related to the description of the
underlying event (UE), via MC tunes, and the settings adopted for the
modelling of colour reconnection (CR), extra initial or final state
QCD radiation (ISR/FSR) and the
description of additional interactions accompanying the hard scatter
(see Ref.~\cite{Seidel} for further details).
Despite some difference in the details of the systematic uncertainty
evaluation, and the use of different baseline MC setups in ATLAS with
respect to CMS, these uncertainty classes are typically assumed to be
fully correlated between measurement from different experiments.

\subsection{Experimental uncertainties}
Experimental uncertainties stem from the modelling of the physics
objects used in the analysis for the event reconstruction and from
the description of the detector response. These are related to the
identification, reconstruction and calibration of leptons, jets, and
missing transverse energy, \met, in the selected events (see Ref.~\cite{Costa}
for further details).  The main contributions to the total
uncertainty of the measurements typically originate from the jet
energy scale (JES) and $b$-tagging related uncertainties.

The various sub-components of the JES systematic uncertainty are
carefully mapped across experiments and their estimated correlation (range)
stems from detailed discussions including experts and analysers from both
the ATLAS and CMS collaborations. While some components are
statistical or detector specific in nature, those stemming from
modelling of the jet (flavour) response, are typically assumed to be
correlated (fully or partially). The exact amount of correlations
across corresponding JES uncertainty categories and their estimate
variation range, useful when evaluating the stability of the
combination results, are described in Ref.~\cite{JESRecom}.

Similarly to the JES, although the algorithm implementations and the
evaluation of the $b$-tagging uncertainties follows different
approaches within ATLAS and CMS, a correlation pattern between
corresponding uncertainty sources has been identified, flagging as
correlated those uncertainty components stemming from general and
method specific physics modelling effects (see Ref.~\cite{BTAGRecom} for
more details).

Additional detector modelling uncertainties, including trigger efficiencies,
uncertainties on the data to MC modelling of the lepton identification,
reconstructions and energy scale, as well uncertainties stemming from
the \met\ and pile-up effects are typically assumed to be uncorrelated
between ATLAS and CMS (this applies also to the JES and $b$-tagging
contributions when sub-dominant).

\subsection{Background related and luminosity uncertainties}
Uncertainties on the background normalisation and shape are in general
analysis, kinematic selection, and final state dependent, but can in turn
affect the measured top-quark properties.  These  uncertainties
are classified into two main categories, based on their origin:
MC-based or data-driven. The former is typically assumed to be fully
correlated across measurements from different experiments. The latter
category is taken as uncorrelated and applies for example to the
data-driven estimate of the contribution of the fake leptons to the
signal and to the normalisation of the $W$+jets background component.

Finally, the uncertainties stemming from the luminosity measurement are
divided into correlated and uncorrelated parts. The correlated part
originates from the common methodologies exploited in the Van der
Meer scan analyses by both collaborations. On the other hand, the
uncorrelated part includes experiment specific effects that could
affect the estimated integrated luminosity of a given data sample (for
example the beam conditions at the specific LHC interaction point, long
term stability of the measurements, and the calibration of the
detectors used for the luminosity measurements).

\section{LHC combination overview}

\begin{table}[tbp!]
\scriptsize
\begin{center}
\begin{tabular}{|l|rcrc|rcrc|} \hline
     Overview                    & 
\multicolumn{4}{|c|}{$\sigma(t\bar t)$  [pb]}    &
\multicolumn{4}{|c|}{$\sigma(t)$ 8 TeV  [pb]}    \\

   (Sept. 2014)         & 
\multicolumn{2}{|c|}{7 TeV }      &  
\multicolumn{2}{|c|}{8 TeV }      &
\multicolumn{2}{|c|}{${\rm t-ch}$} &  
\multicolumn{2}{|c|}{${\rm t}W$}   \\\hline

value	                    & 173.3 &           & 241.4 &             & 85.3 &            & 25.0 &            \\
statistics ${(\star)}$      &   2.8 &(0.08)\ruu &   1.4 & (0.03)\rnu  &  4.1 & (0.11)\rnu &  1.5 & (0.10)\rnu \\
MC model/ theory            &   4.9 &(0.23)\rcc &   4.1 & (0.23)\rnp  &  7.7 & (0.40)\rnp &  4.0 & (0.72)\rnp \\
Detector model ${(\dagger)}$&   4.6 &(0.21)\rcu &   2.7 & (0.10)\rnu  &  5.5 & (0.20)\rnp &  1.2 & (0.06)\rnp \\
JES/Jets ${(\odot)}$        &   2.1 &(0.04)\rcu &   1.7 & (0.04)\rnp  &  4.5 & (0.14)\rnu &  1.3 & (0.08)\rnu \\
Background	            &   2.3 &(0.05)\rpp &   2.3 & (0.07)\rnp  &  3.2 & (0.07)\rnp &  0.6 & (0.02)\rnu \\
Luminosity                  &   6.3 &(0.39)\rcp &   6.2 & (0.53)\rnp  &  3.4 & (0.08)\rnp &  0.7 & (0.02)\rnp \\ \hline
Total uncertainty           &  10.1 &           &   8.5 &             & 12.2 &            &  4.7 &            \\
Relative unc. / Comb. improv. [\%]         &   5.8 &   10.6        &   3.5 &  9.6
& 14.3 &   10.6         & 18.8 &   14.3         \\ 
 \hline
Best single meas.           &   
\multicolumn{2}{|c|}{182.9 $\pm$ 6.3 } & 
\multicolumn{2}{|c|}{242.4 $\pm$ 9.5} & 
\multicolumn{2}{|c|}{83.6 $\pm$ 7.8} & 
\multicolumn{2}{|c|}{27.2 $\pm$ 5.8} \\

\multirow{2}{*}{Ref. ({\tiny\color{blue} ATLAS}, {\tiny\color{red} CMS})} &
\multicolumn{2}{|c|}{\tiny{\color{blue}arXiv}} &
\multicolumn{2}{|c|}{\tiny{\color{blue}arXiv}} &
\multicolumn{2}{|c|}{\tiny{\color{red}JHEP}} & 
\multicolumn{2}{|c|}{\tiny{\color{blue}ATL-CONF}} \\

   &
\multicolumn{2}{|c|}{\tiny{\color{blue}1406.5375}} &
\multicolumn{2}{|c|}{\tiny{\color{blue}1406.5375}} &
\multicolumn{2}{|c|}{\tiny{\color{red}06 (2014) 090}} & 
\multicolumn{2}{|c|}{\tiny{\color{blue}2013-100}} \\\hline

\end{tabular}
\end{center}
%
\scriptsize
\begin{center}
\begin{tabular}{|l|rc|rcrc|rc|} \hline
     Overview                    & 
\multicolumn{2}{|c|}{$m_{\rm top}$ [GeV]}&
\multicolumn{4}{|c|}{$W$ polarisation} &
\multicolumn{2}{|c|}{$A_C$}  \\

   (Sept. 2014)         & 
\multicolumn{2}{|c|}{} &
\multicolumn{2}{|c|}{$F_0$} &
\multicolumn{2}{|c|}{$F_L$} &
\multicolumn{2}{|c|}{} \\  \hline

value	                    &  173.29 &            & 0.626 &            & 0.359 &            & 0.005 &             \\
statistics ${(\star)}$      &    0.24 & (0.06)\ruu & 0.035 & (0.35)\ruu & 0.022 & (0.38)\ruu & 0.007 & (0.61)\rnu  \\
MC model/ theory            &    0.59 & (0.38)\rcc & 0.034 & (0.33)\rcp & 0.019 & (0.30)\rcp & 0.002 & (0.07)\rnp  \\
Detector model ${(\dagger)}$&    0.32 & (0.12)\rcu & 0.020 & (0.11)\rcu & 0.011 & (0.11)\rcu & 0.004 & (0.21)\rnu  \\
JES/Jets ${(\odot)}$        &    0.61 & (0.42)\rcp & 0.020 & (0.11)\rcu & 0.012 & (0.12)\rcu &       &             \\
Background	            &    0.09 & (0.01)\rpp & 0.019 & (0.10)\rcu & 0.010 & (0.09)\rcu & 0.003 & (0.11)\rnp  \\
Luminosity                  &         &            &       &            &       &            &       &             \\ \hline
Total uncertainty           &    0.95 &            & 0.059 &            & 0.035 &            & 0.009 &             \\
Relative unc. / Comb. improv. [\%]         &    0.5  & 10.4    & 9.5  &  22.4       & 9.7   &  23.9          & 181   &  18.2            \\  
\hline

Best single meas.           &   

\multicolumn{2}{|c|}{172.22 $\pm$ 0.73} & 
\multicolumn{2}{|c|}{0.659 $\pm$ 0.027} & 
\multicolumn{2}{|c|}{0.350 $\pm$ 0.026} & 
\multicolumn{2}{|c|}{0.006 $\pm$ 0.011} \\

\multirow{2}{*}{Ref. ({\tiny\color{blue} ATLAS}, {\tiny\color{red} CMS})} &
\multicolumn{2}{|c|}{\tiny{\color{red}CMS-PAS-TOP}} &
\multicolumn{2}{|c|}{\tiny{\color{red}CMS-PAS-TOP}} &
\multicolumn{2}{|c|}{\tiny{\color{red}CMS-PAS-TOP}} &
\multicolumn{2}{|c|}{\tiny{\color{blue}JHEP}} \\ 

   &
\multicolumn{2}{|c|}{\tiny{\color{red}14-001}} &
\multicolumn{2}{|c|}{\tiny{\color{red}13-008}} &
\multicolumn{2}{|c|}{\tiny{\color{red}13-008}} &
\multicolumn{2}{|c|}{\tiny{\color{blue}1402 (2014) 107}} \\ \hline

\end{tabular}
\end{center}
\caption{Summary of the LHC combination results as of September 2014. For each combination,
  the combined result, the total uncertainty ($\sigma_{\rm tot}$) and its breakdown into
  different uncertainty classes ($\sigma_{\rm i}$) is provided 
  [$^{(\star)}$ includes MC statistics and method calibration uncertainties. 
  $^{(\dagger)}$ when not available separately, this uncertainty class
  includes luminosity and JES systematic uncertainties. 
  $^{(\odot)}$ when not available separately, this category includes the jet
  resolution and reconstruction systematics].
  Values
  in brackets are defined as $\sigma_i^2/\sigma_{\rm tot}^2$, and
  quantify the relative importance of each source of uncertainty with
  respect to the total. In addition, the relative precision of the combined result and the
  relative improvement with respect to the most precise
  input measurement are also provided. The last row in the table reports the most precise
  single measurement to date with the corresponding reference (the
  colour code indicates whether the measurements are from the ATLAS or
  CMS collaboration. 
The symbols
  $\circ,~\ast,~\bullet$ describes sources of uncertainties
  with are uncorrelated, partially correlated, or fully
  correlated respectively. Each pair of symbols stands for the
  correlation of measurements from the same experiment or
  across different experiments, respectively. For example \rcu\ indicates
  a source of uncertainty which is fully correlated for
  measurements stemming from the same experiment, but that it
  is assumed to be uncorrelated between ATLAS and CMS ({\it e.g.} the detector modelling uncertainty).  The
  symbol $\times$ is used when only one measurement per
  experiment is used in the combination.
  Cross section total uncertainties are quoted without the beam energy contribution.
}
\label{tab:one}
\end{table}

Several LHC combinations of the top-quark production cross sections
and top-quark properties have been performed in the last few
years. These will be briefly described in the following
subsections. An overview of all results is given in
Table~\ref{tab:one}, together with some details on the total
uncertainty breakdown, on the baseline correlation assumptions for
different uncertainty sources (within and across experiments), as well as
on the most precise single measurement available at the time of the TOP2014
conference.

\subsection{Top-quark production cross sections}

The combination of the top-quark production cross sections (denoted in
the following as $\sigma(\ttbar)$ or $\sigma(t)$) are
performed using LHC data at different centre of mass energies
($\sqrt{s}=7,~8$~TeV), as well as exploiting different production
mechanisms (\ttbar\ pair- and single top-quark production in the $Wt$-
and $t$-channel).

The LHC combination of $\sigma(\ttbar)$ at 7 TeV~\cite{xsec7TeV} uses
as inputs the individual ATLAS and CMS combinations, both featuring
measurements from different $\ttbar$ final states. The result
$\sigma(\ttbar| 7~{\rm TeV}) = 173.3 \pm 10.1$~pb, achieves a relative
precision of 5.8\%, and it is dominated by uncertainties stemming from
the luminosity measurements and MC modelling. The breakdown of the
uncertainties of the combined result, according to statistics and
systematic effects originating from the MC or detector modelling, JES
and jet reconstruction, background modelling and the luminosity is
listed in Table~\ref{tab:one}, together with the references and the
results of the most precise single measurements available at the time
of the TOP2014 conference. Although individual experiment results with
improved precision are already available, the next combination effort
is planned after the completion of the final LHC Run-I results in both
collaborations.
Using $\sqrt{s}=8$~TeV $pp$ data, a combination of the
$\sigma(\ttbar)$ has been performed using as inputs the ATLAS and CMS
cross section measurements from the $e\mu$ dilepton channel~\cite{xsec8TeV}. The result,
$\sigma(\ttbar| 8~{\rm TeV}) = 241.4 \pm 8.5$~pb,
corresponds to a relative uncertainty of 3.5\%, and as in the case of
the corresponding 7 TeV result, it is dominated by uncertainties on
the luminosity measurement and on the MC modelling.

Measurement of the single top-quark production cross
section at $\sqrt{s}=8$~TeV in the $Wt$- and $t$-channel are
available and yielded the results: $\sigma(t-ch| 8~{\rm TeV}) = 85.3
\pm 12.2$~pb and $\sigma(Wt| 8~{\rm TeV}) = 25.0 \pm 4.7$~pb,
respectively~\cite{tchxsec8TeV, wtxsec8TeV}. The combined $Wt$-channel
cross section results is used also to set a lower limit on the CKM
matrix element $V_{tb}$: $|V_{tb}|>0.79$ at 95\% CL. Both combined
cross section results are dominated by systematic uncertainties
originating for the MC modelling. The $t$-channel combination is
based on partial data samples, corresponding to about one fourth of
the available $pp$ collision statistics at $\sqrt{s}=8$~TeV. The planned update
of the combination will profit from newly available measurements with
increased precision, as well as from the ongoing harmonisation efforts
on the treatment of the MC generator uncertainties.

\subsection{Top quark mass}

Several combinations of the top-quark mass (\mt) including
LHC~\cite{massLHC1, massLHC2} and Tevatron~\cite{massWA} results are
available to date (see Refs.~\cite{massWA,Peters} for more details about the
\mt\ world combination).

Despite the availability of updated individual measurements, the LHC \mt\
combinations constituted important milestones for the subsequent
combination of Tevatron and LHC results~\cite{massWA}, and motivated
several topical discussions and harmonisation efforts
(partly still ongoing) within both collaborations. These are aimed at an improved
treatment of the correlation between JES systematic sub-components~\cite{JESRecom},
and at alleviating possible double counting effects between the JES
and MC modelling systematics (in particular as far as uncertainties
from the choice of the hadronisation models are concerned). The latest
combined LHC \mt\ result ($\mt =173.29\pm 0.95$~GeV) has a relative
uncertainty of $0.5\%$, and features individual input measurements in
different \ttbar\ final states, with total uncertainties ranging from
1.06~GeV to 1.63~GeV. The dominant contributors to the total
uncertainty of the combined \mt\ results are the systematic uncertainties
related to the JES, and those stemming from the MC
modelling of the \ttbar\ signal. 

\subsection{$W$ polarisation and charge asymmetry}

Combined results of the $W$ polarisation and of the
charge asymmetry in \ttbar\ events are available and
described in Refs.~\cite{wpol} and \cite{chasym}, respectively.
The $W$ helicity fractions ($F_0$, $F_L$ and $F_R$) from measurements
in different \ttbar\ final states, are combined using a
multi-parameter BLUE implementation in which the correlation between
the $F_L$ and $F_R$, are taken into account ($F_R = 1-F_L-F_0$). The
results, $F_0=0.626\pm 0.059$ and $F_L = 0.359 \pm 0.025$, correspond
to relative total uncertainties below 10\%, and are affected similarly
from the statistical uncertainties and from systematic
uncertainties related to the MC modelling. The combined $W$ helicity
fractions have been used to constrain anomalous couplings ($g_R,~g_L$)
affecting the $Wtb$ vertex in the framework of new physics models. As
in the case of other LHC combinations described in this review,
individual results with improved total uncertainties have recently
become available and will be included in future combinations.

The combination of the charge asymmetry measurements from ATLAS and
CMS features top-quark based asymmetry results from the \ttbar\
lepton plus jets final state, corrected for detector and acceptance
effects. The input measurements as well as the combined result ($A_C =
0.005\pm 0.009$) are currently dominated by statistical
uncertainties.  Due to the large statistical components of the input measurements, the combination improvement amounts to about 20\% for both top-quark properties measurements.


\section{Comments and Conclusions}
In general, statistically limited measurements ({\it e.g.} $W$
polarisation and charge asymmetry) are characterised by the largest
gain in combination precision, and they are largely unaffected by
variation of the assumptions on the systematic uncertainty
correlations.  
On the other hand, systematics dominated measurements, ({\it e.g.} \mt\ and
measurements of the \ttbar\ production cross section), typically
present challenging combinations, and might be significantly affected
by variations of the baseline correlation assumptions across the
different uncertainty sources.  At the same time, they offer a
great opportunity to foster further harmonisation efforts and trigger
refinement of the MC modelling uncertainties as well as of specific
aspects of the MC simulations. In this context, they offer the
largest gain in understanding the complementarity and differences
between measurements and approaches adopted by different experiments.

The TOPLHCWG has been very successful in the past few years. Several
results have been obtained and made public, including the combination of the
production cross sections (\ttbar\ pairs and single top-quark production in the $Wt$-
and $t$-channel), \mt, $W$ helicity and \ttbar\ charge asymmetry
measurements. These have been accompanied by various sets of
recommendations on specific systematic uncertainty splittings and their
correlations between the ATLAS and CMS experiments.

New combinations efforts ({\it e.g.} including additional top-quark properties
and differential cross section measurements) and updates of the
presented combined results are ongoing and will further
improve our knowledge of the top-quark physics sector in the coming years.
These are expected to profit from the progress obtained in the
categorisation of the JES and $b$-tagging systematics across
experiments, as well as from the ongoing harmonisation efforts on
the treatment and evaluation of the MC modelling systematic uncertainties.


\end{document}